\documentclass{iaus}
\usepackage{graphicx}
\usepackage{natbib}

\newcommand\aap{{A\&A}}

\newcommand\apj{{ApJ}}
\newcommand\apjl{{ApJL}}
\newcommand\araa{{ARAA}}

\newcommand\memsait{{Mem.S.A.It.}}

\newcommand\nat{{Nature}}

\newcommand\pra{{Phys. Rev. A}}
\newcommand\solphys{{Solar Physics}}

\title[Non-LTE and 3D effects on Li, Be and B] 
{The light elements in the light of 3D and non-LTE effects}

\author[Martin Asplund \& Karin Lind]   
{Martin Asplund$^1$
 \and Karin Lind$^2$}

\affiliation{$^1$Max-Planck-Institut f\"ur Astrophysik, 
Postfach 1317, D-85741 Garching, Germany \\
email: {\tt asplund@mpa-garching.mpg.de} \\[\affilskip]
$^2$European Southern Observatory, 
Karl-Schwarzschild-Strasse 2, D-85748 Garching, Germany \\
email: {\tt klind@eso.org }}

\pubyear{2010}
\volume{268}  
\jname{Light elements in the Universe}
\editors{C. Charbonnel, M. Tosi, F. Primas \& C. Chiappini, eds.}

\begin{document}

\maketitle

\begin{abstract}
In this review we discuss possible systematic errors 
inherent in classical 1D LTE abundance analyses of late-type stars for the light elements 
(here: H, He, Li, Be and B).
The advent of realistic 3D hydrodynamical model atmospheres
and the availability of non-LTE line formation codes
place the stellar analyses on a much firmer footing and 
indeed drastically modify the astrophysical interpretations in many cases, especially at low metallicities.
For the $T_{\rm eff}$-sensitive hydrogen lines 
both stellar granulation and non-LTE are likely important but the combination of the two
has not yet been fully explored.
A fortuitous near-cancellation of significant but opposite 3D and non-LTE effects leaves the derived
$^7$Li abundances largely unaffected but new atomic collisional data should be taken into account.
We also discuss the impact on 3D non-LTE line formation 
on the estimated lithium isotopic abundances in halo stars in light of recent claims
that convective line asymmetries can mimic the presence of $^6$Li.
While Be only have relatively minor non-LTE abundance corrections, B is sensitive 
even if the latest calculations imply smaller non-LTE effects than previously thought.
\end{abstract}

              

\section{Stellar model atmospheres and spectral line formation}

Stellar chemical compositions are not observed:
to decipher the spectral fingerprints in terms of abundances
requires realistic models for the stellar atmospheres and the
line formation processes. 
Traditionally abundance analyses of late-type stars have been carried out
relying on 1D, time-independent, hydrostatic model atmospheres,
which treat convection with the rudimentary mixing length theory.
All of these are dubious approximations as even a casual glance at the solar 
atmosphere will immediately reveal. The solar atmosphere, as for other late-type stars,
is dominated by granulation, which is the observational manifestation of convection:
an evolving pattern of broad, warm upflows in the midst of narrow cool downdrafts. 
Because of the great temperature sensitivity of the opacity, the temperature drops
precipitously as the ascending gas nears the optical surface before it overturns and
is accelerated downwards. The temperature contrast is therefore very pronounced
in the photospheric layers, amounting to $>1000$\,K at the optical surface
for the Sun (even when averaged over surfaces of equal optical depths these rms-differences
amount to $\approx 400$\,K). In addition to ignoring such atmospheric inhomogeneities,
1D model atmospheres can not be expected to have the correct mean temperature 
stratification because of the simplified convection treatment and the neglect of convective overshoot.
Even small temperature differences can propagate to very large changes in the 
emergent stellar spectrum because of the non-linearities in the radiative transfer and 
the extreme opacity variations. 

Over the past decade or so, 3D, time-dependent, hydrodynamical model atmospheres
for a range of stellar parameters have started to be developed and applied to stellar abundance work
\citep[e.g.][and references therein]{2005ARA&A..43..481A}.
Such 3D models solve the standard hydrodynamical conservation equations coupled
with a simultaneous solution of the 3D radiative transfer equation and therefore self-consistently
predict the convective and radiative energy transport
\citep[see e.g.][for further details]{2009LRSP....6....2N}. 
To make the 3D modelling computationally tractable, the radiative transfer is not solved
for the many thousands of wavelength points as routinely done in 1D model atmosphere codes. 
Instead the frequencies are sorted in opacity and more recently also in 
wavelength space for $\approx 4-20$ bins for which the radiative transfer is solved.
The resulting total radiative heating/cooling as a function of atmospheric depth
is surprisingly well reproduced. 
The 3D models are based on 
similarly realistic microphysics (equation-of-state and continuous/line opacities) as employed
in standard 1D model atmospheres. 
Currently there are mainly four different codes being used to develop 3D models --
{\sc stagger} \citep[e.g.][]{2009LRSP....6....2N}, 
{\sc co5bold} \citep[e.g.][]{2009MmSAI..80..711L}, 
{\sc muram}  \citep[e.g.][]{2005A&A...429..335V} and 
{\sc antares}  \citep[e.g.][]{2009arXiv0905.0177M} --
although essentially all abundance related work to date has been performed within the first two 
collaborations. 

\begin{figure}[t]
\begin{center}
\includegraphics[width=11cm,angle=0]{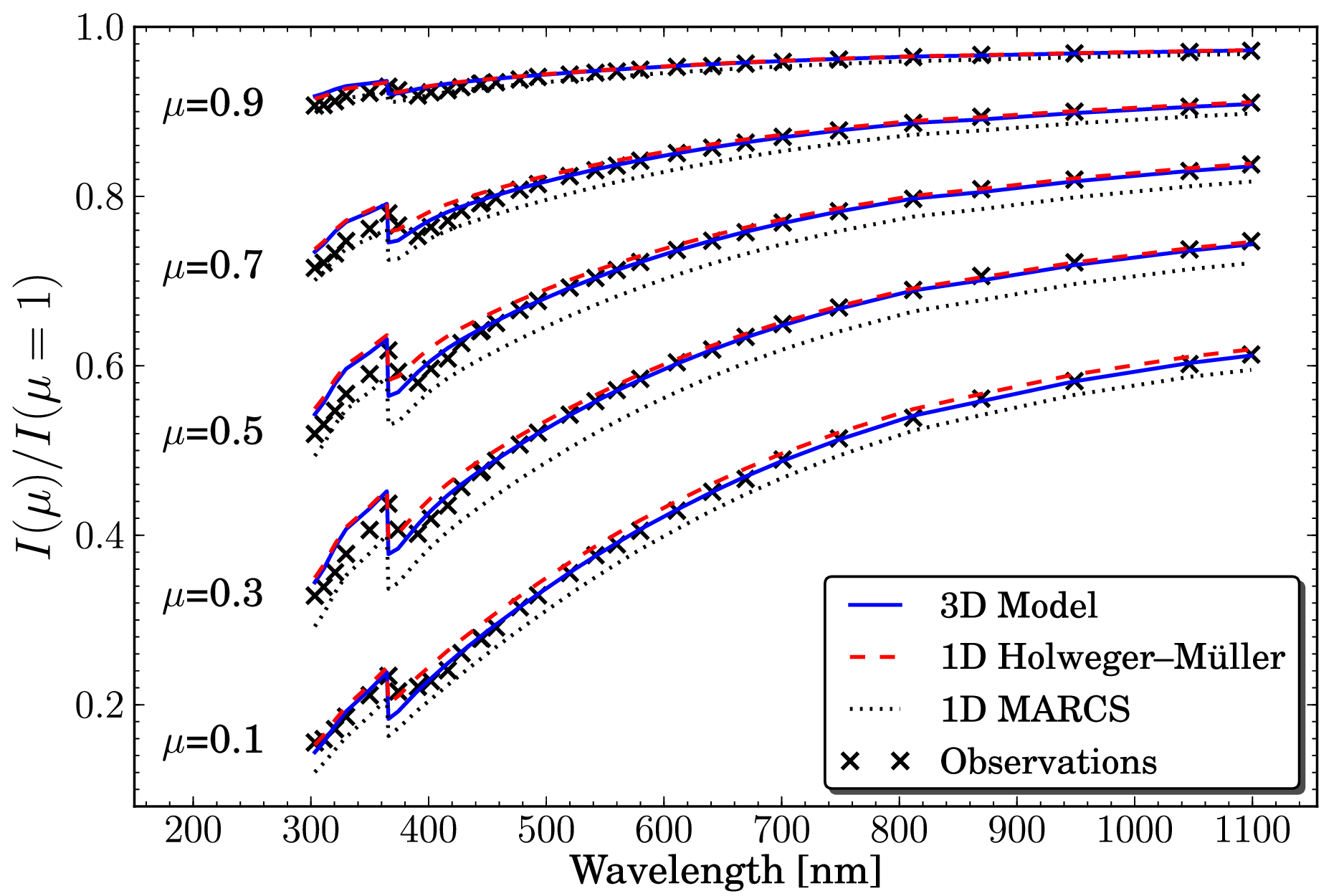}
\caption{Comparison of the observed continuum center-to-limb variation as a function
of wavelength \citep{1994SoPh..153...91N} 
against the predictions for different solar model atmospheres.
The 3D solar model employed by \citet{2009ARA&A..47..481A}
outperforms the 1D theoretical {\sc marcs} model and even
the 1D semi-empirical  \cite{1974SoPh...39...19H} model atmosphere.
}
\label{f:clv}
\end{center}
\end{figure}

Being more sophisticated in the modelling does not automatically translate to being more
realistic. Over the past few years, substantial effort has therefore been dedicated to
verify the suitability of the 3D models for quantitive stellar spectroscopy using an arsenal of observational
diagnostics. Some of the striking successes are that the 
3D models accurately predict the detailed solar granulation properties 
\citep[e.g.][]{2009LRSP....6....2N} as well as spectral line profiles, including their asymmetries and shifts
\citep[e.g.][]{2000A&A...359..729A}, which strongly suggests that the 3D modelling captures 
the essence of the real atmospheric structure and macroscopic gas motions. 
Another crucial test is  the continuum center-to-limb variation, which is an excellent probe
of the mean temperature stratification in the solar atmosphere.
As shown in Fig. \ref{f:clv}, the latest generation of 3D models computed with the {\sc stagger} code
reproduces the observations  extremely well
\citep{Pereira10}; the {\sc co5bold} solar model does a similarly good job (H.-G. Ludwig, private communication).
This is a remarkable achievement. 
The 3D solar model not only outperforms all tested theoretical 1D model
atmospheres like the Kurucz, {\sc marcs} and {\sc phoenix} flavours in this respect, it also does noticeably 
better than the \cite{1974SoPh...39...19H} semi-empirical model, which was constructed largely to fulfill
this observational constraint. 
The 3D solar model also performs very well when confronted with other tests, such 
as the spectral energy distribution, H lines  \citep{Pereira10}  and spatially resolved line profiles 
\citep{2009A&A...508.1403P, 2009A&A...507..417P}.
In all aspects the most recent 3D solar models are clearly highly realistic and can therefore safely
be trusted for abundance analysis purposes
\citep{2009ARA&A..47..481A}.

The success in the solar case gives some credence to the 3D modelling 
being similarly realistic also for other stars for which far fewer indisputable 
observational tests are available.
3D models now exist for a sizable portion of the HR-diagram covering
essentially spectral types A, F, G, K, M and even later 
\citep[e.g.][]{2007A&A...469..687C, 2009MmSAI..80..711L}. 
Both dwarfs, subgiants, giants and even some supergiants have been simulated
and for a wide range of metallicities.
Qualitatively the granulation behaves similarly to the Sun for most of these stars
but typically the convection becomes more vigorous towards higher $T_{\rm eff}$ 
and lower $\log g$.
Perhaps the most marked difference compared with 1D models are
apparent at low [Fe/H], where 3D models
have much cooler temperatures in the optically thin regime 
\citep{1999A&A...346L..17A}. 
This is a natural consequence of the 
dominance of adiabatic heating over radiative heating in absence of a
heavy line-blanketing at low metallicities.
The low temperatures are bound to affect especially minority species, low excitation lines
and molecules, with 3D LTE abundance corrections often amounting to $>+0.5$\,dex
\citep[e.g.][]{2005ARA&A..43..481A}.
One should be aware though that the steep temperature gradients may significantly 
enhance the non-LTE effects compared to the 1D case.

With a stellar model atmosphere it is possible to compute the emergent spectrum, 
which normally for late-type stars is done within
the LTE approximation and in 1D. It should not come as a surprise that this approach often leads
to severe systematic errors.
In recent years, efforts towards rectifying these shortcomings have mainly
focused on 1D non-LTE investigations and 3D LTE calculations,
but the combination of 3D and non-LTE is still largely unexplored
\citep[e.g.][and references therein]{2005ARA&A..43..481A}.
When solving the non-LTE problem one must solve the rate equations for all relevant
levels involving both radiative and collisional transitions coupled with a simultaneous
solution of the radiative transfer equation for all necessary frequencies. 
Needless to say, it rapidly becomes far more cumbersome than in LTE 
with a great deal of additional atomic data that must be considered. 
While the radiative processes like transition probabilities and photo-ionization cross-sections
are now in reasonably good shape -- especially for the non-complicated cases like the 
light elements -- the main uncertainty in non-LTE calculations often stems from
the poorly known collisional excitation and ionization due to electrons and H as well
as charge transfer reactions. Especially the inelastic H collisions are a major problem in
stellar abundance analyses.
Mostly simple but likely erroneous classical recipes like the 
\citet{1968ZPhy..211..404D} formula are applied, often with an additional ad-hoc scaling factor. 
For a few elements, Li being a notable case in point, detailed quantum mechanical
or laboratory measurements exist
\citep[e.g.][]{2003PhRvA..68f2703B}, which reveal that the \citet{1968ZPhy..211..404D} formula 
typically overestimates the real cross-sections by several orders of magnitude.
An alternative approach is to attempt to calibrate the unknown collisional rates using
various observations, such as obtaining consistent results from different lines
or using center-to-limb variations.
Not surprisingly, such empirical procedures suggest a range of scaling factors, 
from $\le 0.1$ for 
Mn \citep{2008A&A...492..823B},
to $\approx 1$ for O \citep{2009A&A...508.1403P} or even $>1$ for Fe \citep{2003A&A...407..691K}. 
Whether these are realistic values, 
or merely artificial results due to too simplified modelling (e.g. 1D) remains to be seen.
It is also worth remembering that a common scaling factor for all transitions
within the same species, let alone between different elements, is highly unlikely.
In spite of the remaining uncertainties introduced by our still poor handle on 
all atomic data, the alternative of relying on the LTE 
approximation will most often lead to even more severe systematic errors. 


\section{The light elements}

\subsection{Hydrogen}

As the most abundant element and the main provider of continuous opacity, it is 
not possible to derive the stellar H abundance spectroscopically.
The H lines are still of great use
in stellar physics as a probe of the atmospheric conditions. In late-type stars 
the wings of the Balmer lines reflect the effective temperature and the temperature gradient
near the continuum forming layers with only a relatively small gravity dependence.
The great temperature sensitivity arises from the high excitation potential of the lower
level ($\chi_{\rm exc}=10.2$\,eV) of the Balmer lines. 
Fortunately, both the Stark broadening and self-broadening are now well established 
\citep[e.g.][]{2000A&A...363.1091B,2008A&A...480..581A}; the new broadening
results in significantly lower $T_{\rm eff}$, especially at low [Fe/H].

The H lines are however sensitive to the convection treatment, which in 1D normally
is estimated by the mixing length theory (MLT) and its inherent four free parameters. 
Before comparing the best-fitting $\alpha_{\rm MLT}$ and $T_{\rm eff}$ between
different studies it is therefore important to bear in mind the various flavours of MLT existing
in different 1D model atmospheres. 
Given the rudimentary nature of MLT to describe the correct convective energy transport
and overshoot near/in the optical surface, 1D-based  $T_{\rm eff}$-scales from H lines
will always be uncertain in an absolute sense even if highly accurate relative values can
be obtained 
\citep[e.g.][]{2006ApJ...644..229A}.
As explained above, 3D hydrodynamical models do provide an attractive alternative 
through their self-consistent computation
of the convection without the need to invoke free parameters. 
\cite{2009A&A...502L...1L} have investigated the 3D LTE formation of H$\alpha$, H$\beta$
and $H\gamma$ lines for six different 3D models, including the Sun and metal-poor dwarfs/subgiants
and compared with 1D models computed with identical microphysics but with different
MLT implementations; see also \citet{Sbordone10} for further discussions.
\cite{2009A&A...502L...1L} found a very complex dependence on the 3D corrections
to the 1D-based estimates of $T_{\rm eff}$ with differences amounting to $\pm 300$\,K
depending on the line in consideration and the $T_{\rm eff}$, [Fe/H] and $\alpha_{\rm MLT}$ of the
1D model. An extra complication arises from the different line shapes in 
3D and 1D, which can not be directly translated to a $T_{\rm eff}$ difference
without specifying exactly which wavelength regions have been used for the comparison.
The prospect for using pre-tabulated 3D corrections to 1D-based $T_{\rm eff}$ estimates
is therefore not encouraging. 
It would be advantageous to directly compare observations with 3D predictions,
which should be possible in the near future with the advent of grids of 
3D models. 

Until recently it has always been argued that the wings of the H Balmer wings are
formed in LTE. \cite{2007A&A...466..327B} has investigated whether this is indeed
correct and found that it is not necessarily so. 
In particular he found that electron collisions are not sufficient
to establish LTE. In terms of $T_{\rm eff}$, the non-LTE calculations would imply
$\ge 100$\,K higher values than in LTE. This however depends crucially on
the still uncertain inelastic H collisions -- improved atomic physics calculations
are clearly needed to resolve this issue. 
It would also be necessary to perform 3D non-LTE H line computations, especially
at low metallicity. Such calculations have recently been performed in the solar case
with encouraging results \citep{Pereira10}.

\subsection{Helium}

While in hot stars the very highly excited transitions of He\,{\sc i} enable
a determination of the He abundance (see Kaufer, these proceedings), in late-type stars
the corresponding atomic levels are not sufficiently populated at typical photospheric temperatures.
Instead the He\,{\sc i} lines (e.g. 1083.0\,nm) have a chromospheric origin and can neither
probe the He abundance nor the photospheric temperature structure.

\subsection{Lithium}

\begin{figure}[t]
\begin{center}
\includegraphics[width=10cm,angle=90]{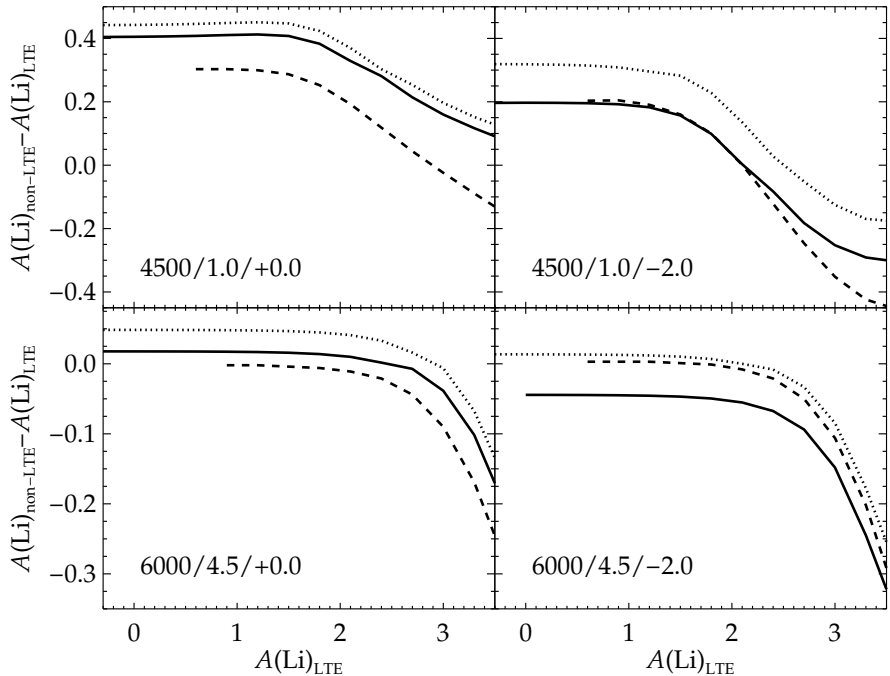}
 \caption{The 1D non-LTE abundance corrections for a selection of stellar parameters
 (labelled with $T_{\rm eff}/\log g/$[Fe/H]) from the study of \cite{2009A&A...503..541L}. 
 The solid line denotes the case with all transitions considered while the dotted line
 correspond to the case when the charge transfer reactions of \citet{2003A&A...409L...1B}
 are ignored. Also shown as dashed lines are the non-LTE corrections from 
 \cite{1994A&A...288..860C}. }
\label{f:Li_NLTE}
\end{center}
\end{figure}

Stellar Li abundance are usually determined only from the
resonance line at 670.7\,nm but in exceptional cases also from the weaker
subordinate line at 610.4\,nm. Contrary to most elements, the simplicity of
the Li atom has enabled quantum mechanical calculations of all atomic data
necessary for non-LTE investigations, especially of cross-sections for
collisional excitation and charge transfer with neutral hydrogen 
\citep{2003A&A...409L...1B}. 
%
The widely used study of
\cite{1994A&A...288..860C} mapped out the non-LTE effects in
late-type stars and their detailed variation with metallicity, effective
temperature, surface gravity, and lithium line strength. Recently,
\cite{2009A&A...503..541L} revisited the 1D non-LTE line formation 
with improved collisional data and treatment of
line-blocking (see Fig \ref{f:Li_NLTE}). The investigations have shown that two competing
effects govern the population levels of Li. Over-ionization from the first
excited level of Li\,{\sc i} is prominent at low effective temperatures
due to a strong $J_\nu-B_\nu$ excess bluewards of the photo-ionization
threshold at 350\,nm. When the Li line is weak, the balance between UV
over-ionization from the first excited state and over-recombination to
higher excited levels determines the population levels of Li\,{\sc i}. The
resulting non-LTE corrections are typically minor, smaller than
$\pm0.1\rm\,dex$, but increasing up to
$+0.5$\,dex in red giants. However, when the line is close to saturation,
the loss of line photons establishes an efficient recombination and
de-excitation ladder, which increases the population of low-excited levels of neutral Li. 
Furthermore, the scattering-dominated resonance line source function drops far below 
$B_\nu$. The non-LTE corrections then change sign and reach $-0.5$\,dex in extreme cases.
\cite{2009A&A...503..541L} have made available convenient routines to
interpolate the non-LTE corrections for a wide range of
stellar parameters and Li abundances. 

In LTE the Li\,{\sc i} level populations depend on the local conditions, especially the
gas temperature. 
The much cooler mean temperature stratifications in 3D models 
than corresponding 1D models at low metallicity
translate to a 3D LTE abundance correction of $\approx -0.3$\,dex for the Li\,{\sc i}
670.8\,nm resonance line in metal-poor dwarfs and subgiants
\citep[e.g.][]{1999A&A...346L..17A}. 
In addition, the presence of atmospheric inhomogeneities typically also increase the
line strength. 
One should be aware however that the steeper temperature gradients 
in the 3D models are likely to boost the over-ionization.
Spatially resolved
observations of the solar surface show how the line actually is weaker in
the bright, granular upflows, despite the drastic drop in temperature that
they undergo. This can be understood since the intense UV radiation field
in the granules produces pronounced over-ionization \citep{1997ApJ...489L.107K}.

\cite{2003A&A...399L..31A}  performed full
3D non-LTE computations for the Sun and two metal-poor stars using
a 21-level Li model atom. They concluded
that the  line strengthening due to the cooler
temperatures in the upper atmospheric layers on the one hand and line weakening
from increased over-ionization on the other hand largely cancel each other.
\cite{2003A&A...409L...1B} updated these results in light of new collisional data,
especially charge transfer reactions and reached similar conclusions. 
The 3D non-LTE calculations thus agree with the 1D non-LTE results to within
0.05--0.1\,dex. 
In fact, even better agreement between the two is seen in
the recent study of extremely metal-poor stars by \cite{Sbordone10} 
with difference of $<0.03$\,dex, at least compared with the particular family
of 1D model atmospheres they used. 
\cite{Sbordone10} used a smaller atom (8 levels) but computed for more 3D 
models. They also provided a very handy functional relation between the
Li abundance and the line strength, which allows direct 
computation of the 3D non-LTE based abundance without invoking any
particular 1D model atmosphere. 
The \cite{1982A&A...115..357S} Li-plateau as inferred from 1D analyses 
is thus unlikely to be seriously in danger 
although the exact slope with for example [Fe/H] may be affected
\citep[e.g.][]{2006ApJ...644..229A,Sbordone10},
which deserves to be studied closer with improved 3D non-LTE calculations.

\subsection{$^6$Li/$^7$Li}

The minor isotope $^6$Li can be detected through the isotopic shift in the 
Li\,{\sc i} 670.8\,nm line. 
The distortion of the line profile is very small and therefore necessitates extremely high-quality spectra
($S/N>400$, $\lambda/\Delta\lambda > 100,000$ and minimal fringing).
At solar metallicity, possible blending lines must also be carefully evaluated
\citep{2003A&A...405..753I}, a problem which however disappears at low [Fe/H].
The first positive detection in a halo star was claimed by  \cite{1993ApJ...408..262S} for HD\,84937. 
\cite{2006ApJ...644..229A} boosted the number of $>2\sigma$
detections to 10 (including HD\,84937) using high-quality VLT/UVES spectra
analysed both in 1D and 3D LTE.
The derived $^6$Li abundance at [Fe/H]$<-2.5$ can not be easily accommodated
with standard Galactic cosmic ray production, which has spurred a number of studies of more 
speculative production channels, including non-standard Big Bang nucleosynthesis
with supersymmetric particles 
\citep[e.g.][]{2009NJPh...11j5028J}.

Several potential problems linger over the published $^6$Li/$^7$Li analyses. 
\cite{2009A&A...504..213G} have suggested that residual fringing in the observed spectra
can cause large uncertainties. Their Subaru/HDS spectra are, however, much more
inflicted by fringing than the VLT/UVES spectra of \cite{2006ApJ...644..229A}, who furthermore
carefully assessed their possible impact on the results.
More worrisome is the restriction to 1D or 3D LTE line formation. 
The atmospheric convective motions typically result in C-shaped line asymmetries
\citep[e.g.][]{2000A&A...359..729A}, which
can thus mimic the presence of $^6$Li. 
Indeed, \cite{2007A&A...473L..37C} have argued based on 
3D non-LTE calculations that the \cite{2006ApJ...644..229A} results
must be reevaluated. 
\cite{2010arXiv1001.3274S} have considered this in more detail and concluded
that the derived $^6$Li/$^7$Li ratios in a 1D analysis 
are overestimated by typically 0.015, which if true could explain the 
tendency for most stars with non-significant detections in the \cite{2006ApJ...644..229A} 
sample to cluster around $^6$Li/$^7$Li$\approx 0.01$.
Furthermore, \cite{2010arXiv1001.3274S} argue that by taking this convective effect
into account, only 4 out of 10 stars would remain as $>2\sigma$ detections. 

We argue that the \cite{2010arXiv1001.3274S} results should not be over-interpreted,
since  there are subtle but crucial differences in the analyses. 
\cite{2010arXiv1001.3274S} rely only on the Li line itself to determine all
parameters: $^7$Li and $^6$Li abundances, wavelength shift and intrinsic line broadening
(rotation in 3D but in 1D also macro-/microturbelence) while  \cite{2006ApJ...644..229A} 
rely on a range of Fe, Ca etc lines to first determine the line broadening which is then 
fixed for the Li profile fitting. 
Since the other lines are also asymmetric, this is accounted for in the profile fitting
through a slightly larger required broadening, which then compensates for the use of
symmetric Li lines in 1D.  \cite{2006ApJ...644..229A}  also did a 3D LTE analysis for all lines
and found very similar results as in 1D but with much improved profile fits. 
We have also performed 3D non-LTE line calculations for Li similar to \cite{2010arXiv1001.3274S}
and can in fact exactly reproduce their results and their claimed 0.015 effect in
$^6$Li/$^7$Li when only using the Li line and not any other lines. 
We argue though that this does not make full use of the
information encoded in the spectra due to the degeneracy between the four fitting parameters. 
Indeed, their procedure is akin to determining the D abundance in QSO absorption system
only from the Ly$\alpha$ line without first resolving the intrinsic velocity structure of the 
H\,{\sc i} clouds from other metallic lines. 
Similarly problematic is to rely on 3D non-LTE for Li but 3D LTE for the other lines,
in view of the likely pronounced non-LTE effects. 
The only way forward to convincingly demonstrate whether or not $^6$Li is present
in detectable amounts in the atmospheres of halo turn-off stars is to perform
full 3D non-LTE calculations for all lines considered. 
We are currently working on this and will report the results elsewhere.

In summary, it is not yet possible to say that $^6$Li has definitely been detected but it is 
definitely too early to say that $^6$Li has not been detected.
We also note that even  \cite{2010arXiv1001.3274S} agree that $^6$Li is present
in HD\,84937 and a few other stars. Since $^6$Li is always destroyed and more so than
$^7$Li, one would still end up with a very significant cosmological $^6$Li problem when
invoking stellar depletion to solve the cosmological $^7$Li problem 
\cite[e.g.][]{2006ApJ...644..229A,2006Natur.442..657K, 2009A&A...503..545L}.

\subsection{Beryllium}

In late-type stars the Be abundance can in practice only be derived from the
resonance doublet of Be\,{\sc ii} at 313\,nm. 
In contrast to Li and B, Be is normally present in significant amounts in both neutral and singly ionized form
due to its high ionization potential, which makes Be less prone to over-ionization. 
The  Be\,{\sc ii} line formation proceeds under non-LTE conditions but because 
the two dominant non-LTE effects -- over-ionization and over-excitation -- largely compensate
each other the resulting non-LTE abundance corrections are typically minor
\citep{1995A&A...297..787G}. 
Both the lower and upper level of the doublet are over-populated relative to the
LTE  prediction.  
Be\,{\sc i} tends to be somewhat over-ionized due to a UV radiation excess ($J_\nu/B_\nu > 1$), which 
simultaneously produce an over-excitation of the upper level of the Be\,{\sc ii} 313\,nm transitions.
\cite{GarciaPerez10} have investigated the non-LTE line formation
of the Be\,{\sc ii} 313\,nm doublet across a wide range of stellar parameters. 
They found that at solar metallicity the two effects almost perfectly balance each other 
with the resulting non-LTE abundance corrections amounting to $|\Delta \log \epsilon_{\rm Be}| \le  0.05$\,dex. 
At low metallicity the non-LTE corrections are positive ($\Delta \log \epsilon_{\rm Be} \approx 0.1$\,dex) 
for $T_{\rm eff} \ge 6000$\,K and negative  ($\Delta \log \epsilon_{\rm Be} \approx -0.1$\,dex) for 
lower $T_{\rm eff}$.
While the non-LTE effects will not qualitatively change the conclusions for
the metallicity evolution of Be (Primas and Boesgaard, these proceedings), accounting for 
the new non-LTE calculations will tend to make the slope somewhat shallower. 

The 3D line formation of the Be\,{\sc ii} lines have so far only been investigated 
for the Sun
\citep{2009ARA&A..47..481A}.
The Be\,{\sc ii} lines are not very sensitive to 3D effects  at least 
within the LTE approximation. The 3D non-LTE case has not yet been investigated
but given the relatively modest non-LTE effects in 1D one would not expect a great difference
in 3D, although this prediction should be verified with detailed calculations.

\subsection{Boron}

As for Be, both over-ionization and over-excitation are at play in the formation of the 
B\,{\sc i} resonance lines at 249.7\,nm and 209.0\,nm but because B\,{\sc i} is the minority
ionization stage for $T_{\rm eff} \ge 6000$\,K 
the effects on the resulting line profiles are much more dramatic \citep{1996A&A...311..680K}. 
The over-ionization is largely driven by photo-ionization from the excited 
B\,{\sc i} levels since $J_\nu/B_\nu > 1$ at the relevant wavelengths. 
It is also assisted by pumping in the B\,{\sc i} resonance lines, which
produces a sufficient over-population of these upper levels. 
The same pumping results in a substantial increase in the line source function 
$S_\nu/B_\nu \approx J_\nu/B_\nu > 1$ for the 249.7 and 209.0\,nm lines. Thus, in contrast to the case of Be,
for the B\,{\sc i} resonance lines the two non-LTE effects work in tandem to decrease 
the non-LTE line strengths substantially compared with the LTE prediction. 

\begin{figure}[t]
\begin{center}
\includegraphics[width=11cm,angle=0]{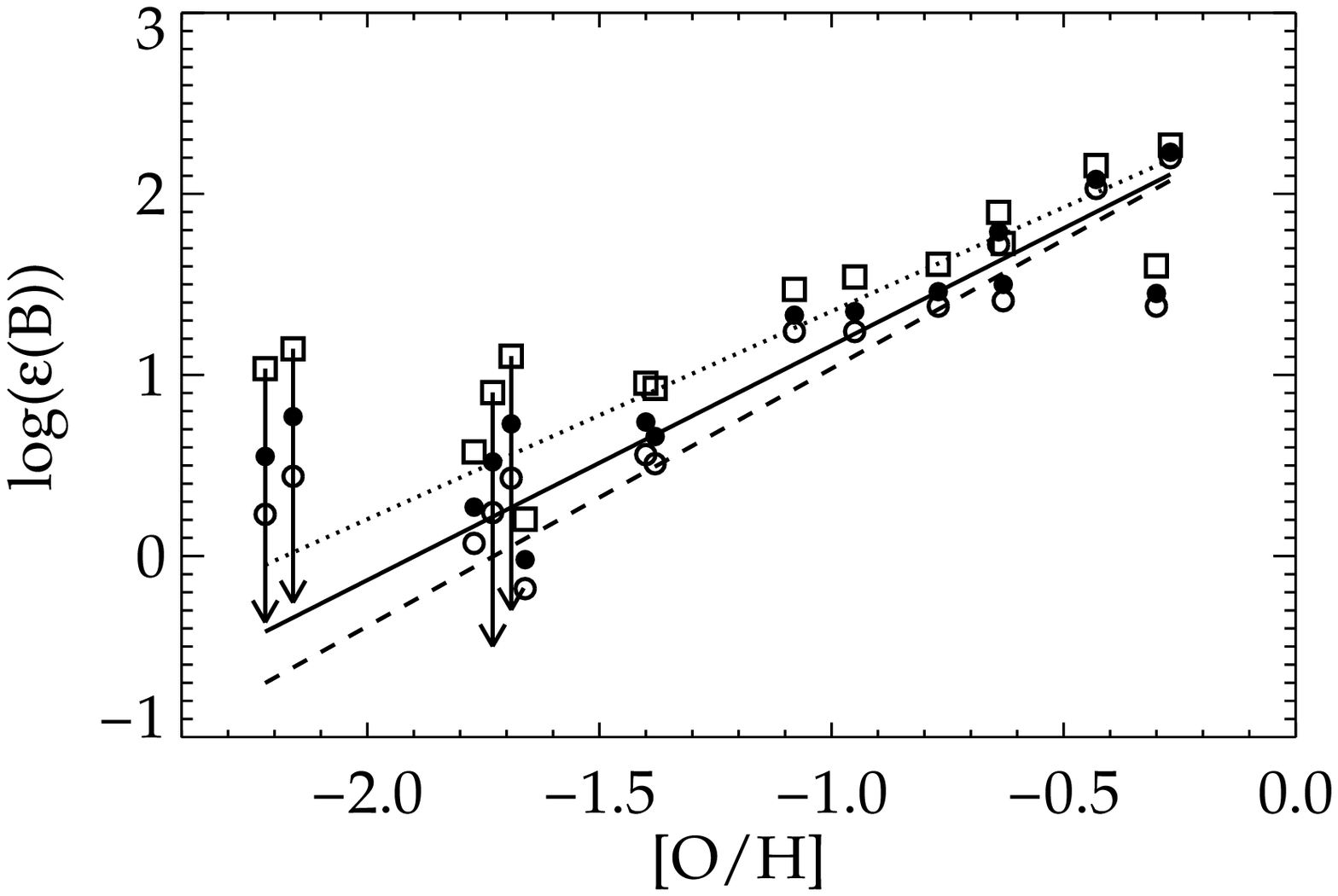}
\caption{The derived stellar B abundances in halo stars \citep{Tan10} as a function of 
[O/H] for
LTE (open circles with dashed line the least square fit) 
and non-LTE (filled circles and solid line); 
upper limits are marked with arrows. The O abundances
have been corrected for non-LTE effects according to 
\citet{2009A&A...500.1221F}.
Also shown are the LTE-based B abundances of \citet{Tan10} but corrected according to
the non-LTE calculations of \citet{1996A&A...311..680K} (open squares and dotted line).
Inelastic H collisions have not been included for the B non-LTE results but for O according
to the \citet{1968ZPhy..211..404D} formula.
}
\label{f:B}
\end{center}
\end{figure}

\citet{1996A&A...311..680K} computed non-LTE abundance corrections for the
B\,{\sc i} lines for a grid of {\sc marcs} 1D model atmospheres. 
The non-LTE effects grow in size towards higher $T_{\rm eff}$ and lower [Fe/H].
Indeed, they found non-LTE abundance corrections as large as $>+0.5$ for typical turn-off halo stars. 
As both non-LTE processes feed on the UV radiation field, it is important to include
background line opacities both for the calculations of the photo-ionization rates and the
resonance lines, which \citet{1996A&A...311..680K} did in a somewhat approximate manner.
Recently,  \citet{Tan10} confirmed the importance
of over-ionization and pumping but they obtained markedly less severe non-LTE effects, which
can be traced to a more complete treatment 
of the line-blocking. 
Their estimated non-LTE abundance corrections amount to $\approx +0.3$\,dex at the lowest [Fe/H],
when not including the highly uncertain inelastic H collisions according to 
\citet{1968ZPhy..211..404D}; the corrections would be $\approx +0.1$\,dex
when this formula would be applied without any further scaling factor.
As shown in Fig. \ref{f:B}, the new non-LTE results of \citet{Tan10} make the
resulting evolution of B as a function of metallicity somewhat steeper with 
a slope of $\approx 1.3$.
The reader is referred to Primas (these proceedings) for a discussion of the 
astrophysical implications of such a correlation.

With the exception of the Sun \citep{2009ARA&A..47..481A},
the 3D line formation of the B\,{\sc i} lines has not been investigated, neither in
LTE nor in non-LTE. In view of the prominent non-LTE effects and the general
expectation that they would become even more pronounced in 3D due to
the atmospheric inhomogeneities, performing detailed 3D non-LTE calculations
for B\,{\sc i} should have a high priority.

\section{Concluding remarks}

Important progress has been made over the last decade
in terms of improving the abundance analyses of the light elements in late-type stars by
accounting for non-LTE as well as 3D effects. 
The non-LTE line formation of the resonance lines of Li\,I, Be\,II, and B\,I share common properties, 
but the final non-LTE abundance corrections are always element specific. 
Especially differences in ionization potential and the extent to which UV radiative transitions 
contribute to the excitation and ionization balance govern the LTE departures.
There are however still several outstanding analysis problems. 

Signs are that the new generation of 3D hydrodynamical model atmospheres are indeed
highly realistic and thus trustworthy for abundance purposes as they successfully reproduce 
most, if not all, observational tests they have been exposed to. The key development for the future will
be to make such 3D models generally available for a wide range of stellar parameters and
actually to get stellar abundance practitioners 
to start using them routinely. The main obstacle here will likely be to overcome
old habits rather than say lack of computational power, since it is now possible to perform a 3D-based 
solar/stellar abundance analysis for thousands of lines, at least in LTE 
\citep{2009ARA&A..47..481A}.
On the non-LTE side, the challenges are two-fold: improving the necessary atomic data, especially
for collisions, and to couple the non-LTE line formation
with 3D model atmospheres, which is still very much unchartered territory. 
This is now certainly feasible but will require additional (wo)manpower.
Only then can we with some degree of confidence argue that our analysis efforts match
the investments in obtaining impressive observational data. 
As outlined above, such modelling efforts will no doubt lead to many reinterpretations of
observations with profound implications for our understanding of stellar, galactic and cosmic evolution
in general and the light elements in particular. 






\end{document}